# *SAMBLASTER*: fast duplicate marking and structural variant read extraction

Gregory G. Faust[1] and Ira M. Hall[1,2,*]

[1]Department of Biochemistry and Molecular Genetics, University of Virginia, Charlottesville, VA, USA

[2]Center for Public Health Genomics, University of Virginia, Charlottesville, VA, USA

**ABSTRACT**

**Motivation:** Illumina DNA sequencing is now the predominant source of raw genomic data, and data volumes are growing rapidly. Bioinformatic analysis pipelines are having trouble keeping pace. A common bottleneck in such pipelines is the requirement to read, write, sort and compress large BAM files multiple times.

**Results:** We present SAMBLASTER, a tool that reduces the number of times such costly operations are performed. SAMBLASTER is designed to mark duplicates in read-sorted SAM files as a piped post-pass on DNA aligner output before it is compressed to BAM. In addition, it can simultaneously output into separate files the discordant read-pairs and/or split-read mappings used for structural variant calling. As an alignment post-pass, its own runtime overhead is negligible, while dramatically reducing overall pipeline complexity and runtime. As a stand-alone duplicate marking tool, it performs significantly better than PICARD or SAMBAMBA in terms of both speed and memory usage, while achieving nearly identical results.

**Availability:** SAMBLASTER is open source C++ code and freely available from https://github.com/GregoryFaust/samblaster

**Contact:** imh4y@virginia.edu

## 1 FEATURES AND METHODS

The ongoing rapid cost reduction of Illumina paired-end sequencing has resulted in the increasingly common use of this technology for a wide range of genomic studies, and the volume of such data is growing exponentially. Generating high quality variant calls from raw sequence data requires numerous data processing steps using multiple tools in complex pipelines. Typically, the first step in this analysis is the alignment of the sequence data to a reference genome, followed by the removal of duplicate read-pairs that arise as artifacts either during PCR amplification or sequencing. This is an important pipeline step, as failure to remove duplicate measurements can result in biased downstream analyses.

### 1.1 Common usage scenario: piped SAM input

Extant duplicate marking programs such as PICARD *MarkDuplicates* (http://picard.sourceforge.net/) and SAMBAMBA *markup* (https://github.com/lomereiter/sambamba) require position-sorted SAM or BAM (Li, et al.) as input, and perform two passes over the input data, thereby requiring that their input file be stored on disk. Instead, SAMBLASTER marks duplicates in a single pass over a read-id sorted SAM file. This allows the SAM output of alignment tools such as NOVOALIGN (http://www.novocraft.com) or BWA-MEM (http://arxiv.org/abs/1303.3997) to be piped directly into SAMBLASTER, which marks duplicates and outputs read-id sorted SAM, which in turn is piped to SAMTOOLS or SAMBAMBA for sorting and/or compression, without the need to store any intermediate files. This saves one compression-write-read step in the common case in which a duplicate marked position-sorted file is needed later in the pipeline, and two such cycles if a duplicate marked read-sorted file is also needed. The elimination of each such cycle is a significant cost savings of both disk space and runtime. For example, using ~50X-coverage whole genome sequence data for NA12878 from the Illumina Platinum Genomes (ENA Accession: ERP001960), each compressed BAM file consumes over 100 GB of space and requires 7+ hours to compress with SAMTOOLS, and 8.5 hours of CPU time in 1.5 hours elapsed time with SAMBAMBA using 10+ threads on a server class machine. An advantage of the two-pass duplicate marking strategy is that one can retain the "best" read-pair of a set of duplicates, while SAMBLASTER always keeps the first such pair found in its input. However, in practice we find this has negligible impact on variant detection (data not shown).

SAMBLASTER will mark as duplicate any secondary alignments associated with a duplicate primary alignment, and thus works particularly well with BWA-MEM output. As of the time of writing, neither SAMBAMBA nor PICARD has this functionality.

### 1.2 Extracting reads for structural variation detection

Structural Variation (SV) is a major source of genome diversity but is difficult to detect relative to other forms of variation. SV detection algorithms typically predict SV breakpoints based on the distribution of discordant paired-end alignments, in which the paired reads map to either end of an SV breakpoint, and/or split-read alignments where reads aligns across an SV breakpoint. Many SV detection algorithms require long runtimes due to the overhead associated with searching for and extracting these alignments from large BAM files comprised predominantly of uninformative read-pairs. However, some SV detection algorithms, including HYDRA (Quinlan, et al.) and LUMPY (http://arxiv.org/abs/1210.2342), are able to input files comprised solely of discordant and/or split-read mappings, which are typically >100-fold smaller in size. This presents an opportunity to greatly increase pipeline efficiency by extracting discordant and split-read mappings during a prior pipeline step that already requires reading through the entire dataset. SAMBLASTER is able to extract such reads directly from the SAM output of an aligner such as BWA-MEM that can detect both discordant read-pairs and split read mappings. In addition, when used with other popular paired-end aligner such as BWA-ALN or NOVOALIGN which do not identify split-read mappings, SAMBLASTER can extract unmapped and clipped reads for realignment with a sensitive split-read alignment tool such as YAHA (Faust and Hall) for later use to detect SV. By including these capabilities directly in a tool that also marks duplicates, several SV detection pipeline steps can be eliminated.

---

[*]To whom correspondence should be addressed.

### 1.3 Custom data structures

SAMBLASTER uses a custom data structure that uses significantly less memory than competing duplicate marking programs. It considers two or more read-pairs to be duplicates when they have the same *signature*, defined as the combination of the sequence, strand, and reference position of both reads in the pair. To most accurately define the reference positions, it parses the CIGAR string to calculate where the 5' end of each read would align to the reference genome under the assumption that the entire read is mapped. This is similar to the strategy used by PICARD. To detect duplicates, it builds a set of such signatures, marking a read-pair as duplicate if its signature has previously appeared in the input.

To avoid storing a structure with all of this information, the signature is broken into pieces. Each unique combination of sequence1, strand1, sequence2 and strand2 maps to its own position in an array in which a set of the associated position pairs is stored as a hash table. The hash tables are optimized to store 64-bit integers; 32 bits for each reference position. SAMBLASTER thus has low memory requirements relative to other tools, ~20 bytes per read pair, which frees it from the need to use temporary intermediate files. See Figure 1 for details. In addition, SAMBLASTER does not allocate/free any per-read memory structures for reading/writing SAM records, thereby increasing I/O throughput.

### 1.4 Performance evaluation

To evaluate the speed, memory and disk usage of SAMBLASTER as a stand-alone duplicate marking algorithm vs. PICARD *Mark-Duplicates* and SAMBAMBA *markdup*, we used the NA12878 dataset aligned via BWA-MEM as our input source. All timings were performed on a server class machine with 128 GB of RAM and two 8-core (16 thread) Intel Xeon E5-2670 processors running at 2.6GHz. To make the comparison of SAMBLASTER to PICARD as similar as possible, we ran both using SAM for both the input and output format. SAMBAMBA *markdup* does not support SAM format for either input or output. To make the test as comparable as possible, we used uncompressed BAM for both, even though such files are still much smaller than SAM. While SAMBLASTER is single threaded, to show best possible PICARD and SAMBAMBA runtimes, each were allocated 32 threads, and SAMBAMBA single threaded statistics are also shown. The results of the comparison test are shown in Table 1. SAMBLASTER outperforms the other duplicate marking programs in terms of CPU seconds, wall time, disk IO, and memory usage.

**Table 1.** Comparative runtime, memory usage, and disk usage statistics for SAMBLASTER 0.1.14, PICARD *MarkDuplicates* 1.99 and SAMBAMBA *markdup* 0.4.4 as stand-alone duplicate marking tools, and in a common pipeline that produces a duplicate marked position-sorted BAM file as its final output. In the pipeline, SAMBAMBA sort and compression are used. There is also a control pipeline run without duplicate marking which demonstrates that SAMBLASTER adds little overhead. SAMBAMBA *markdup* times are shown for both an uncompressed (ucmp) and compressed (cmp) position-sorted intermediate file. These tests were run using local RAID storage with fast read/write times. In a more common scenario using networked disk access, SAMBLASTER's reduced IO results in greater runtime savings vs. the other tools.

|  | Mark Dups Threads | Extra Disk (GB) | Total Disk IO (G ops) | CPU Time (sec) | Wall Time (min) | Mem Usage (GB) |
|---|---|---|---|---|---|---|
| Stand Alone Mark Duplicates Function ||||||||
| SAMBLASTER | 1 | - | 1.863 | 2077 | 43 | ~15 |
| SAMBAMBA | 1 | - | 2.285 | 6338 | 75 | ~32 |
| SAMBAMBA | 32 | - | 2.285 | 6603 | 54 | ~43 |
| PICARD | 32 | - | 3.056 | 63160 | 302 | ~30 |
| Mark Duplicates – Sort – Compress Pipeline ||||||||
| No duplicate marking | - | - | 1.954 | 51819 | 117 | ~19 |
| SAMBLASTER | 1 | 0 | 1.987 | 52767 | 118 | ~23 |
| SAMBAMBA cmp | 32 | 108 | 2.455 | 86512 | 154 | ~43 |
| SAMBAMBA ucmp | 32 | 391 | 3.997 | 61321 | 163 | ~43 |

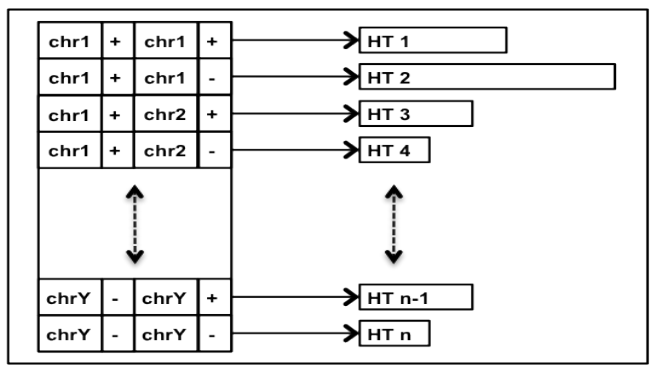

**Figure 1.** Custom data structure in SAMBLASTER with a separate set of reference-offset pairs, stored as a hash table, for each combination of sequence1, strand1, sequence2, and strand2. The hash tables are optimized to store 64-bit integers.

### ACKNOWLEDGEMENTS

We thank Ryan Layer, Colby Chiang, Michael Lindberg, Eric Faust and Aaron Quinlan for their thoughtful insights.

*Funding*: This research was supported by an NIH New Innovator Award DP2OD006493-01 and a Burroughs Wellcome Fund Career Award to IH.